\pdfoutput=1
\documentclass[sigconf]{acmart}

\AtBeginDocument{%
  }

\copyrightyear{2026}
\acmYear{2026}
\setcopyright{cc}
\setcctype{by}
\acmConference[SA Technical Communications '26]{SIGGRAPH Asia 2026 Technical Communications}{December 01--04, 2026}{Kuala Lumpur, Malaysia}
\acmBooktitle{SIGGRAPH Asia 2026 Technical Communications (SA Technical Communications '26), December 01--04, 2026, Kuala Lumpur, Malaysia}
\acmDOI{10.1145/3829339.3847837}
\acmISBN{979-8-4007-2841-9/2026/12}

\usepackage{subcaption}

\definecolor{codebg}{RGB}{248,249,251}
\definecolor{codeaccent}{RGB}{180,190,210}
\definecolor{codekeyword}{RGB}{175,50,115}
\definecolor{codebuiltin}{RGB}{50,115,180}

\begin{document}

\title[Stochastic Graph Compression for Differentiable Light Tracing]{Stochastic Graph Compression for Constant-Memory Differentiable Light Tracing}

\author{Linas Beresna}
\orcid{0009-0004-7102-324X}
\affiliation{%
  \institution{Simon Fraser University}
  \city{Burnaby}
  \country{Canada}}
\email{linas\_beresna@sfu.ca}

\author{Eugene Fiume}
\orcid{0000-0001-5939-7055}
\affiliation{%
  \institution{Simon Fraser University}
  \city{Burnaby}
  \country{Canada}}
\email{eugene\_fiume@sfu.ca}

\renewcommand{\shortauthors}{Beresna and Fiume}

\begin{abstract}
We describe \emph{Reservoir Light Replay Backpropagation} (ResLRB), a method for reverse-mode differentiable light tracing whose memory is constant in path length and in the number of sensor connections per light path. Naive automatic differentiation of a light tracer records a computation graph that grows with the number of valid sensor connections along each path, since flux can be splatted from every scattering vertex; adjoint path replay removes the analogous dependence on depth for viewpoint path tracing, but its non-branching structure does not extend to the splatting case. We compress the adjoint graph during the primal pass by stochastically retaining a single representative sensor connection per light path via a streaming weighted reservoir. The adjoint pass reconstructs the path deterministically by replaying its pseudorandom number sequence and backpropagates only through the retained connection, yielding an unbiased gradient estimator whose peak memory is that of differentiating a single scattering event.
\end{abstract}

\ccsdesc[500]{Computing methodologies~Ray tracing}

\keywords{differentiable rendering, adjoint method, light tracing, constant memory, reservoir sampling}

\begin{teaserfigure}
  \centering
  \includegraphics[width=\linewidth]{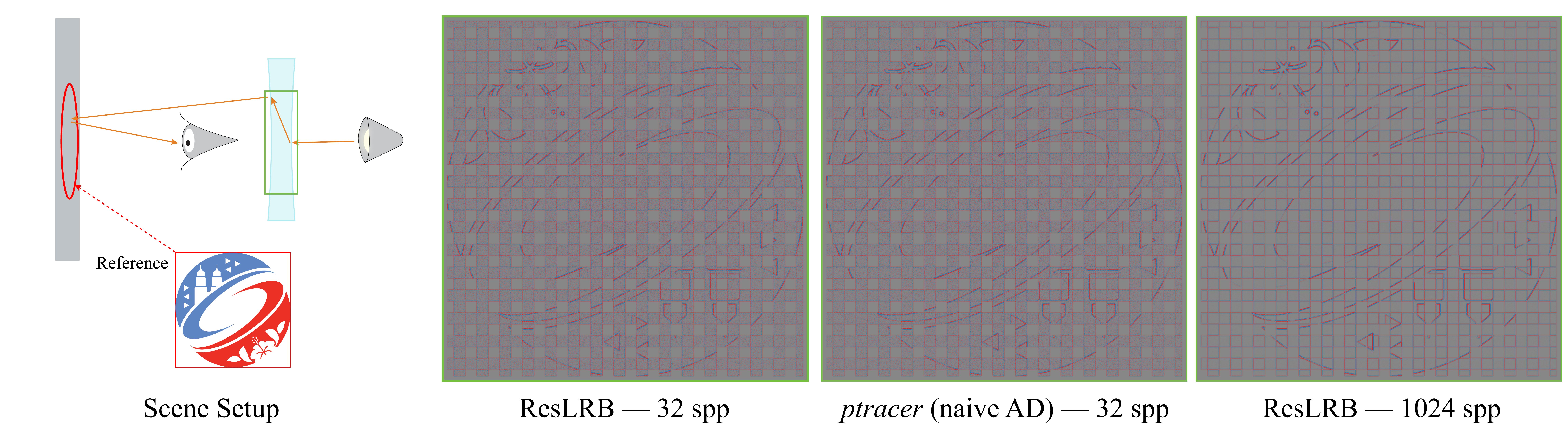}
  \caption{Constant-memory differentiable light tracing. A caustic formed by
light refracting through a glass slab onto a diffuse receiver (left, with
the reference target), differentiated with the attached formulation of
Section~\ref{sec:attached}. Backpropagating an image loss to the slab
geometry, ResLRB and naive AD (\texttt{ptracer}) produce matching gradients
at 32~spp. Beyond 32~spp naive AD exhausts GPU memory, whereas ResLRB holds
memory constant and reaches 1024~spp, yielding the same gradient with less
Monte Carlo noise.}
  \Description{Four side-by-side panels. The first two show a caustic cast by
light refracting through a glass slab onto a diffuse receiver: the optimized
result and the reference target, which closely match. The remaining panels show
gradient images of the slab geometry computed by naive automatic
differentiation at 32 samples per pixel and by ResLRB at 1024 samples per
pixel; the two share the same structure, but the ResLRB gradient is markedly
less noisy.}
  \label{fig:teaser}
\end{teaserfigure}

\maketitle

\section{Introduction}
Differentiable rendering computes the derivatives of a simulated image with respect to scene parameters such as materials, emitters, and geometry, enabling gradient-based optimization for inverse problems. Because the parameter domain is large, these derivatives are computed in reverse mode: reverse-mode automatic differentiation (AD) records every operation of the forward simulation into a computation graph and traverses it backwards, propagating the gradient of a scalar loss to all parameters at once. In a Monte Carlo light transport simulation the graph therefore holds every scattering event along every path and grows with path depth; for deep paths or high sample counts its footprint can exceed available GPU memory. Radiative Backpropagation~\cite{NimierDavid2020Radiative} recasts this gradient computation as a transport problem, trading memory for time.

Path Replay Backpropagation~\cite{Vicini2021PathReplay} (PRB) removes that cost for viewpoint path tracing. The primal pass simulates the path forward and stores only the seed of its pseudorandom number generator (PRNG); the adjoint pass re-seeds that PRNG and re-traces the path, so the identical sequence of random numbers regenerates the identical sequence of vertices. We call this \emph{PRNG replay}: instead of saving the path, it calculates each point on the fly and updates gradients step-by-step. The full graph is never held in memory, giving constant memory in path depth at the cost of one additional traversal.

\begin{figure}[t]
    \centering
    \begin{subfigure}[b]{0.42\linewidth}
        \centering
        \includegraphics[width=\linewidth]{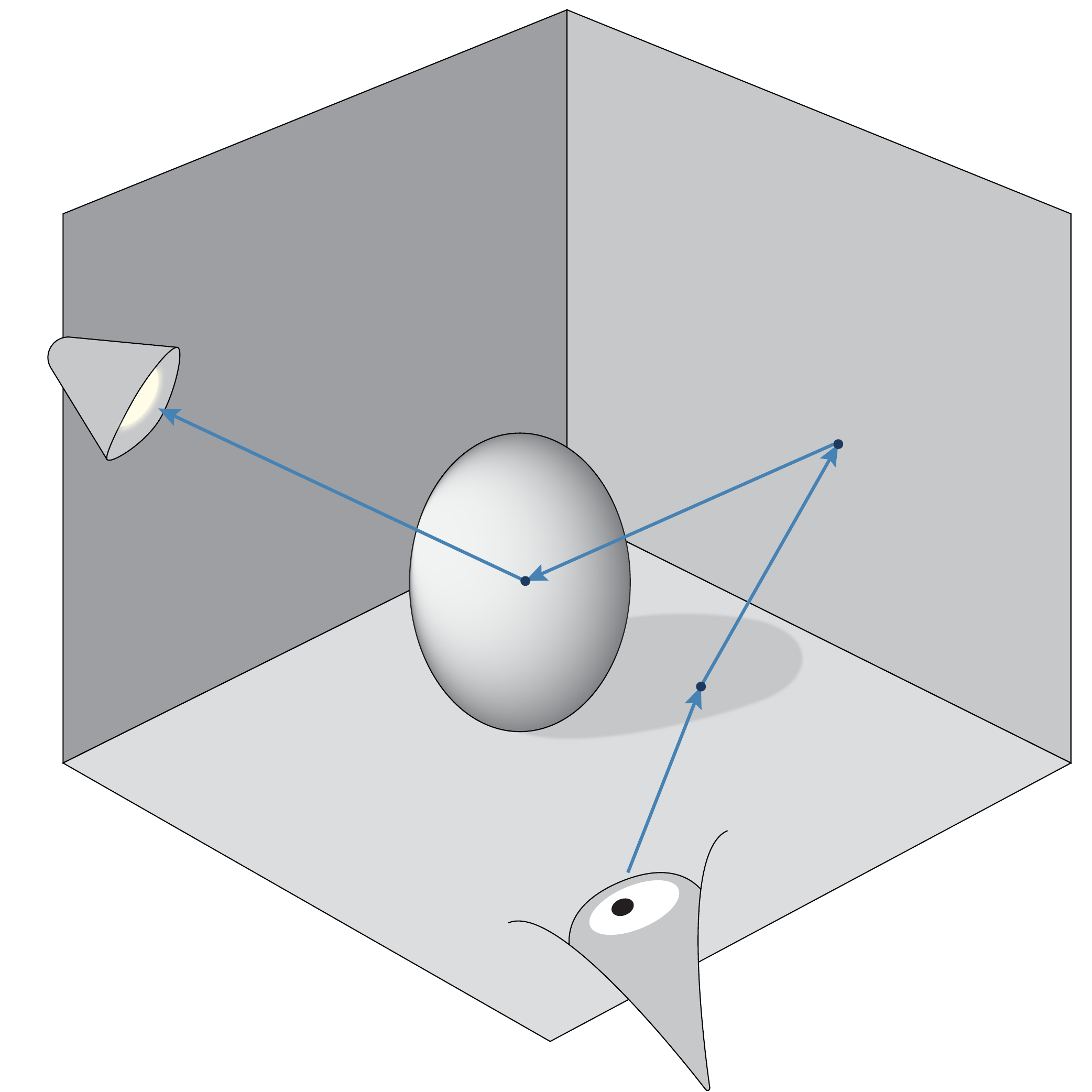}
        \caption{Viewpoint path tracing.}
        \label{fig:path-tracing}
    \end{subfigure}
    \hfill
    \begin{subfigure}[b]{0.42\linewidth}
        \centering
        \includegraphics[width=\linewidth]{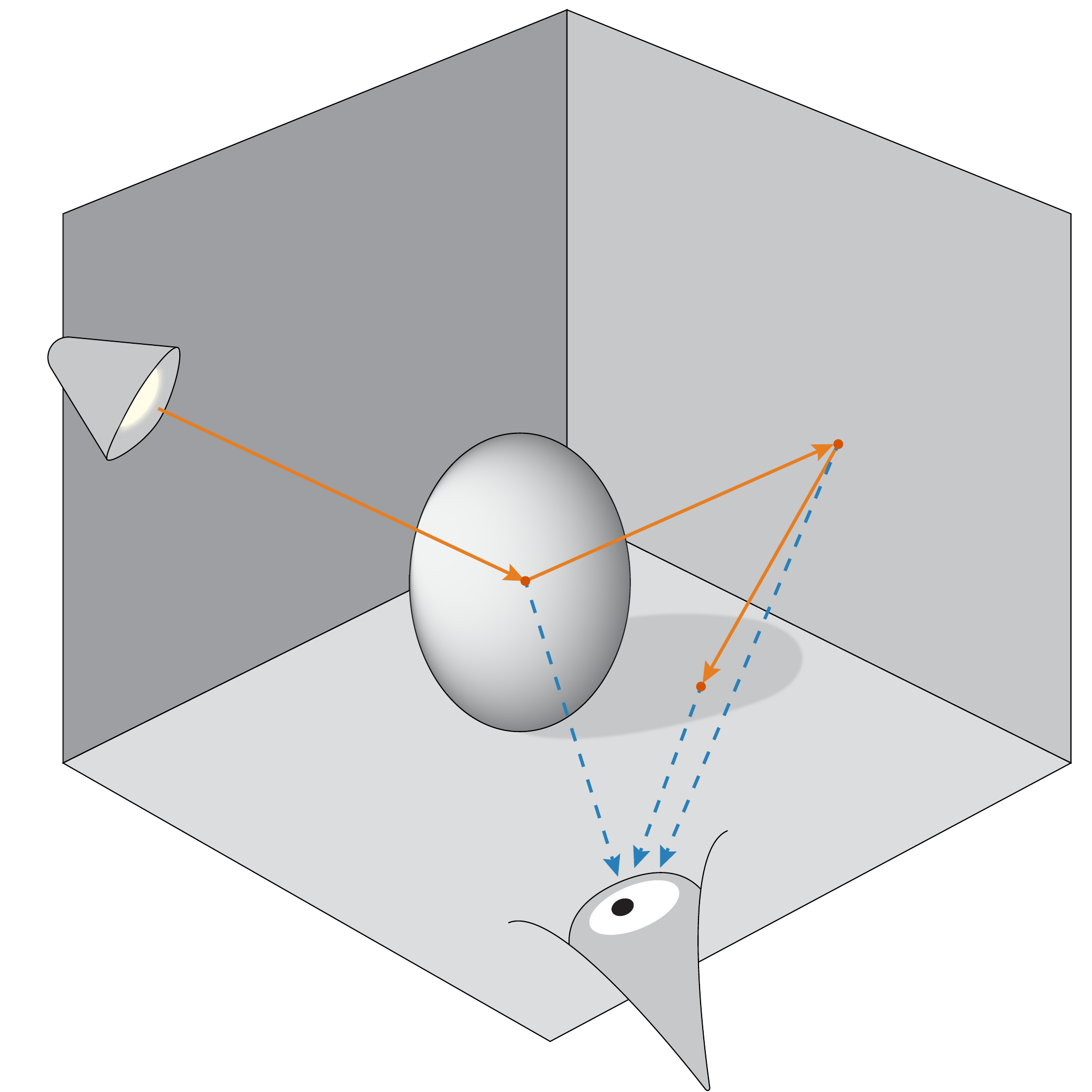}
        \caption{Light tracing.}
        \label{fig:light-tracing}
    \end{subfigure}
    \caption{Path--sensor connectivity. A viewpoint path (left) contributes
    through a single emitter connection, giving a non-branching graph. A
    light path (right) can connect to the sensor at every vertex (dashed
    blue arrows), so its graph grows with both depth and connection count.}
    \Description{Two schematic diagrams of light transport paths. On the left,
a viewpoint path leaves the sensor, scatters off several surfaces, and makes a
single connection to the emitter, forming one unbranched chain. On the right, a
light path leaves the emitter and scatters off the same surfaces, but a dashed
arrow runs from every scattering vertex back to the sensor, so the path
branches into as many contributions as it has vertices.}
    \label{fig:connectivity-comparison}
\end{figure}

This method does not directly apply to light tracing. A viewpoint path typically contributes to the image once, via a single emitter connection, giving a non-branching graph; a light path instead propagates outward from the emitter and can connect to the sensor at every scattering vertex. Hence a path of depth $k$ produces up to $k$ separate contributions, each with its own computation graph (Figure~\ref{fig:connectivity-comparison}). Prior adjoint light tracing has sidestepped this: Lipp et al.~\cite{Lipp2024} target view-independent object-space irradiance. Instead the image-space splatting we consider is what introduces the branching.

We describe a method that obtains reverse-mode gradients for light tracing with memory consumption that is constant in path length and in the number of sensor connections per path, preserving the two-pass structure of PRB. The primal pass retains a single sensor connection per path in a streaming weighted reservoir~\cite{Vitter1985}; the adjoint pass reconstructs the path once by PRNG replay and backpropagates only through the retained connection, yielding an unbiased estimator. Streaming resampling has a long history in rendering for variance reduction and reuse, from Resampled Importance Sampling~\cite{Talbot2005} and ReSTIR~\cite{Bitterli2020} to path-space~\cite{lin2022generalized} and bidirectional variants~\cite{Hedstrom2025}; closest to our work, Nimier-David et al.~\cite{nimierdavid2022unbiased} use reservoir selection in differentiable volume rendering to sample a single distance along a ray. We apply a structurally similar mechanism along a surface light path, but to a different end: bounding the recorded computation graph rather than reducing variance.

\section{Method}
\label{sec:method}

The name \emph{Reservoir Light Replay Backpropagation} (ResLRB) reflects the method's three ingredients: reservoir compression of per-path sensor connections, adjoint light tracing, and path replay~\cite{Vicini2021PathReplay}.

\subsection{Notation}
A light path $\bar{x} = (x_0, \dots, x_k)$ starts from a point $x_0$ sampled on an emitter and propagates outward through the scene. At each scattering vertex $x_i$, $i \geq 1$, the path can be connected to the sensor, contributing flux $C_i$ to a single sensor coordinate $u_i \in \mathbb{R}^2$; the per-path image-space contribution is the multiset $\{(C_i, u_i)\}_{i=1}^k$. Each $C_i$ depends on the differentiable scene parameters $\theta$ through emitter radiance, scattering functions, the geometric term, and sensor importance. We develop the method for the \emph{detached} case, in which $u_i$ is independent of $\theta$ so that the only $\theta$-dependence is through $C_i$, and extend it in Section~\ref{sec:attached} to the \emph{attached} case, where $u_i$ depends on $\theta$ as well.

For a scalar objective $\mathcal{L}$ defined on the rendered image, the gradient $\nabla_\theta \mathcal{L}$ is the quantity we wish to estimate. Writing $\mathcal{L}$ as a sum of per-pixel terms and pulling the per-path contribution out, the gradient associated with a single path is
\begin{equation}
g(\bar{x}; \theta) = \sum_{i=1}^k \nabla_\theta \big[\, f_i(C_i(\theta),\, u_i) \,\big],
\label{eq:per-path-grad}
\end{equation}
where $f_i$ is the pixel-filter-weighted splat of $C_i$ to coordinate $u_i$, treated as linear in $C_i$. Naive reverse-mode AD records one computation graph per term and traverses all $k$ in the backward pass; our goal is to estimate $g(\bar{x}; \theta)$ while recording at most one at a time.

\subsection{Stochastic Selection of One Connection}
\begin{figure}[t]
    \centering
    \includegraphics[width=0.92\linewidth]{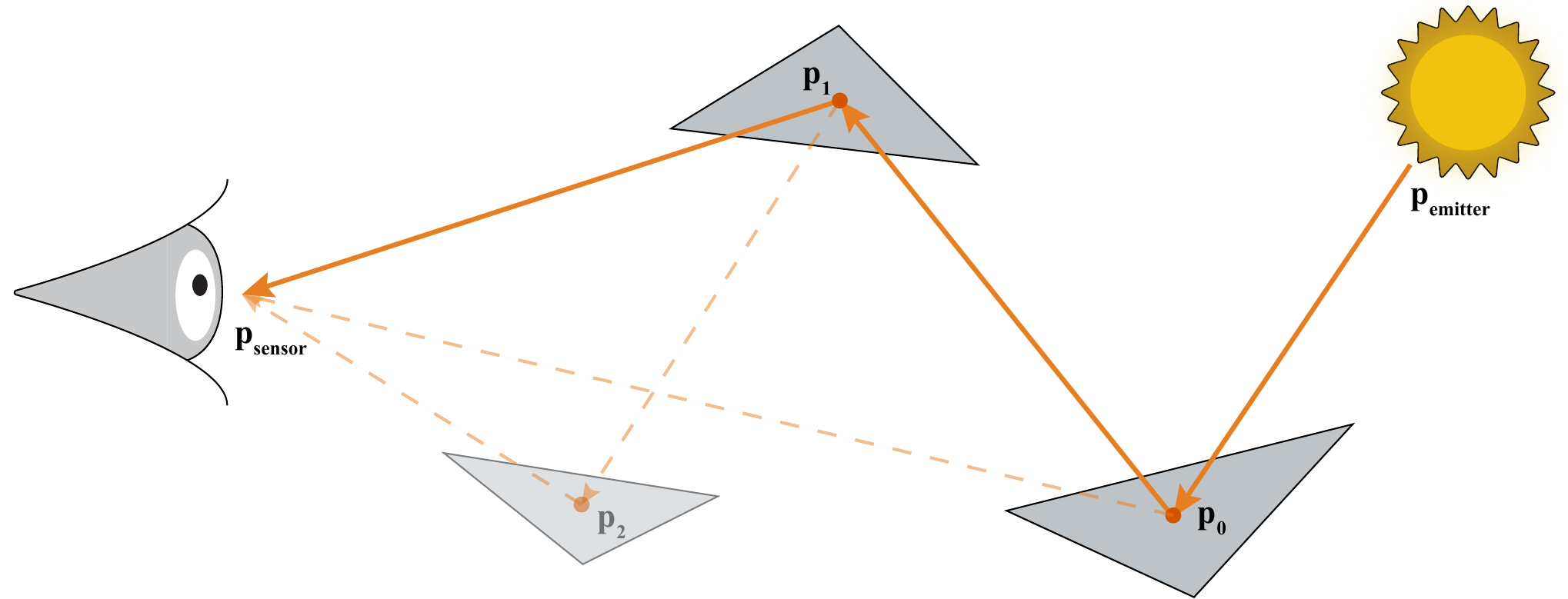}
    \caption{Stochastic selection of a representative connection. Vertices
    $p_0$, $p_1$, $p_2$ each offer a valid sensor connection (dashed). The
    reservoir keeps one (solid, here $p_1$) and discards the rest,
    collapsing the branching adjoint graph into a single path.}
    \Description{A light path from an emitter scatters off three surfaces at
vertices labelled p0, p1 and p2, each offering a dashed candidate connection to
an eye-shaped sensor. Only the connection from p1 is drawn as a solid arrow,
indicating the one candidate the reservoir retained; the connections from p0
and p2 remain dashed and are discarded.}
    \label{fig:reservoir-selection}
\end{figure}

During the primal pass, we stream the candidate connections $\{(C_i, u_i, w_i)\}_{i=1}^k$ into a single-item weighted reservoir, where $w_i = \ell(C_i) \geq 0$ is a scalar weight. Any $\ell$ that is strictly positive wherever a connection contributes leaves the estimator of Section~\ref{sec:unbiased} unbiased, so $\ell$ is free to be chosen for variance-reduction alone. We use luminance: it makes the selection probability proportional to each connection's share of the path's image contribution, concentrating the retained connection on the vertices that dominate the gradient, and it is already computed in the primal pass at no additional cost. The reservoir maintains a running sum $W = \sum_{j=1}^i w_j$ and replaces its current contents with the new candidate at step $i$ with probability $w_i / W$, as illustrated in Figure~\ref{fig:reservoir-selection}. After the path terminates, the reservoir holds index $I \in \{1, \dots, k\}$ selected with probability
\begin{equation}
\Pr(I = i) = \frac{w_i}{\sum_{j=1}^k w_j} = \frac{w_i}{W_k}.
\label{eq:reservoir-pmf}
\end{equation}
This is the standard property of weighted reservoir sampling~\cite{Vitter1985}. Crucially, only the final $W_k$ and the data associated with the selected index $I$ are kept; all other candidates are discarded.

\subsection{Unbiased Estimator}
\label{sec:unbiased}
Given the selected index $I$, we define the random estimator
\begin{equation}
\hat{S}(\bar{x}; \theta) = \frac{W_k}{w_I} \cdot f_I(C_I(\theta),\, u_I).
\label{eq:estimator}
\end{equation}
Taking the expectation over the random selection,
\begin{align}
\mathbb{E}_I\!\left[\hat{S}(\bar{x}; \theta)\right]
&= \sum_{i=1}^k \Pr(I = i) \cdot \frac{W_k}{w_i} \cdot f_i(C_i(\theta),\, u_i) \notag \\
&= \sum_{i=1}^k \frac{w_i}{W_k} \cdot \frac{W_k}{w_i} \cdot f_i(C_i(\theta),\, u_i)
 = \sum_{i=1}^k f_i(C_i(\theta),\, u_i),
\label{eq:unbiased-primal}
\end{align}
which is exactly the deterministic per-path image contribution, so $\hat{S}$ is unbiased for the sum of all $k$ splats. Provided $f_i$ is sufficiently smooth that differentiation and expectation can be interchanged,
\begin{equation}
\mathbb{E}_I\!\left[\nabla_\theta \hat{S}(\bar{x}; \theta)\right]
= \nabla_\theta\, \mathbb{E}_I\!\left[\hat{S}(\bar{x}; \theta)\right]
= g(\bar{x}; \theta),
\label{eq:unbiased-grad}
\end{equation}
so $\nabla_\theta \hat{S}$ is unbiased for the per-path gradient. Variance increases relative to evaluating all $k$ terms, because every sensor connection is not computed, but the estimator remains correct in expectation.

\subsection{Two-Pass Evaluation via Path Replay}
The estimator $\nabla_\theta \hat{S}$ requires the AD graph for only the winning connection, and is computed by PRNG replay~\cite{Vicini2021PathReplay}. In the \emph{primal pass}, we run the light path forward, build the reservoir, and store the PRNG seed together with the selected index $I$ and the final weight sum $W_k$. The splat $f_I$ is evaluated, scaled by $W_k / w_I$, and accumulated into the image; no AD graph is recorded. In the \emph{adjoint pass}, we re-seed the PRNG and re-trace the path up to depth $I$, enabling reverse-mode AD only for the local scattering computation at each vertex while the remaining state is reconstructed in detached form. Each vertex contributes to $\nabla_\theta \hat{S}$ through its local factor in $C_I$: using the cached primal value of $C_I$ as a multiplicative weight, the contribution at vertex $x_i$ is $\nabla_\theta f_s^{(i)} \cdot \big(C_I / f_s^{(i)}\big)$, where $f_s^{(i)}$ is the local BSDF evaluation. The log-derivative identity $\nabla_\theta \log f = \nabla_\theta f / f$ lets each vertex contribute its share of $\nabla_\theta C_I$ without holding any other vertex's graph; the scaling factor $W_k / w_I$ depends only on primal-pass quantities and is treated as a constant.

Each adjoint iteration opens a fresh local AD scope, records the graph for one scattering event, propagates the local gradient into the scene parameters, and discards the graph before the next vertex. Peak AD memory in the adjoint pass is therefore bounded by the cost of differentiating a single scattering event, regardless of $k$ or $I$. Figure~\ref{alg:reslrb} outlines both passes.

\begin{figure}[t]
\setlength{\fboxsep}{6pt}%
\noindent\fcolorbox{codeaccent}{codebg}{%
\begin{minipage}{\dimexpr\columnwidth-2\fboxsep-2\fboxrule\relax}
\footnotesize\ttfamily\raggedright
\textcolor{codekeyword}{\textbf{def}} primal(ray):\\
\hspace*{1em}R, W = \textcolor{codebuiltin}{empty\_reservoir}(), 0\\
\hspace*{1em}v = \textcolor{codebuiltin}{trace}(ray)\\
\hspace*{1em}\textcolor{codekeyword}{\textbf{for}} i \textcolor{codekeyword}{\textbf{in}} \textcolor{codebuiltin}{range}(max\_depth):\\
\hspace*{2em}\textcolor{codekeyword}{\textbf{if not}} v.valid(): \textcolor{codekeyword}{\textbf{break}}\\
\hspace*{2em}C\_i, w\_i = \textcolor{codebuiltin}{connect\_to\_sensor}(v)\\
\hspace*{2em}W += w\_i\\
\hspace*{2em}\textcolor{codekeyword}{\textbf{if}} \textcolor{codebuiltin}{uniform}() < w\_i / W:\\
\hspace*{3em}R.store(v, C\_i, w\_i, i)\\
\hspace*{2em}v = \textcolor{codebuiltin}{trace}(v)\\
\hspace*{1em}\textcolor{codebuiltin}{splat}(R.C * (W / R.w), R.u)\\
\hspace*{1em}\textcolor{codekeyword}{\textbf{return}} R.depth\\[0.5\baselineskip]
\textcolor{codekeyword}{\textbf{def}} adjoint(ray, I, $\delta_L$):\\
\hspace*{1em}v = \textcolor{codebuiltin}{trace}(ray)\\
\hspace*{1em}\textcolor{codekeyword}{\textbf{for}} i \textcolor{codekeyword}{\textbf{in}} \textcolor{codebuiltin}{range}(max\_depth):\\
\hspace*{2em}\textcolor{codekeyword}{\textbf{if not}} v.valid(): \textcolor{codekeyword}{\textbf{break}}\\
\hspace*{2em}\textcolor{codebuiltin}{consume\_reservoir\_rng}()\\
\hspace*{2em}f\_s = \textcolor{codebuiltin}{bsdf\_eval}(v)\\
\hspace*{2em}$\delta_\theta$ += \textcolor{codebuiltin}{backward\_grad}(f\_s, $\delta_L$ * R.C / f\_s)\\
\hspace*{2em}\textcolor{codekeyword}{\textbf{if}} i == I: \textcolor{codekeyword}{\textbf{break}}\\
\hspace*{2em}v = \textcolor{codebuiltin}{trace}(v)
\end{minipage}}
\caption{ResLRB primal (top) and adjoint (bottom) passes.}
\Description{Pseudocode for two procedures. The primal procedure traces a light
path, and at each vertex computes a sensor connection and its weight, adds the
weight to a running sum, and replaces the reservoir contents with probability
equal to the new weight divided by that sum; after the loop it splats the
retained contribution reweighted by the total sum and returns the selected
depth. The adjoint procedure re-traces the same path, consuming the reservoir's
random numbers to stay in step, and at each vertex evaluates the BSDF and
accumulates a parameter gradient weighted by the cached contribution divided by
that BSDF value, stopping once it reaches the selected depth.}
\label{alg:reslrb}
\end{figure}

\begin{figure*}[t]
  \centering
  \setlength{\tabcolsep}{2pt}
  \begin{tabular}{@{}ccc@{}}
    \includegraphics[width=0.30\linewidth]{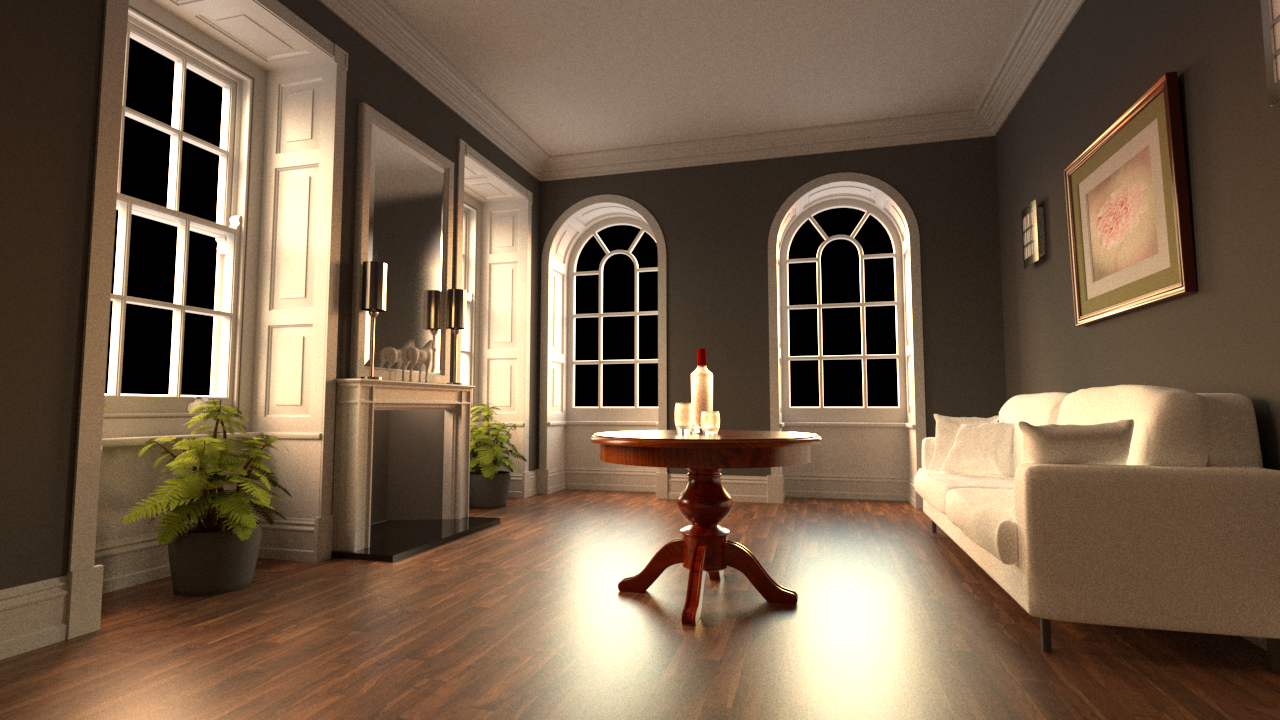} &
    \includegraphics[width=0.30\linewidth]{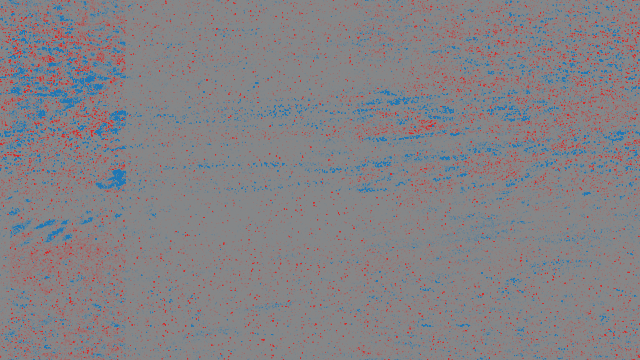} &
    \includegraphics[width=0.30\linewidth]{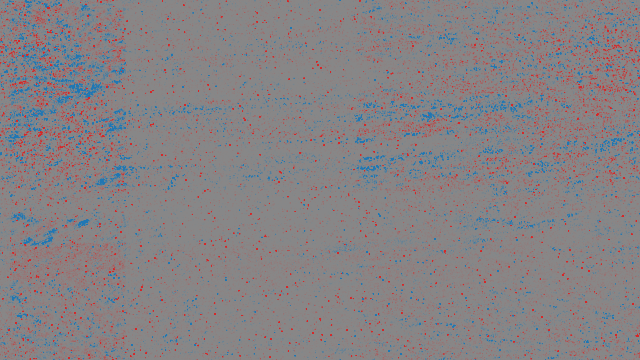} \\
    {\small Reference render} & {\small Naive AD} & {\small ResLRB} \\
  \end{tabular}
  \caption{Gradient validation. Gradient of an $L_2$ image loss with respect
  to the floor's diffuse reflectance in the Grey \& White Room (path length
  32, 32~spp); red is positive, blue negative, grey zero. Naive AD
  (\texttt{ptracer}) and ResLRB agree in sign and spatial structure and
  differ only in Monte Carlo noise, confirming that reservoir selection
  leaves the estimator unbiased.}
  \Description{Three panels. The left panel is a reference render of the Grey
and White Room interior. The middle and right panels are gradient images of the
floor's diffuse reflectance computed by naive automatic differentiation and by
ResLRB respectively, using a red-white-blue colour map for positive, zero and
negative values. Both gradient images show the same red and blue spatial
pattern across the floor and differ only in the amount of speckled Monte Carlo
noise.}
  \label{fig:validation}
\end{figure*}

\subsection{Extension to Attached Differentiation}
\label{sec:attached}
The formulation above is \emph{detached}: the splatted position $u_i$ is
held independent of $\theta$. This fails whenever perturbing $\theta$ moves
the path and shifts $u_i$, most strongly at specular and refractive
vertices, as in the glass-slab caustic of Figure~\ref{fig:teaser}. We extend
to this \emph{attached} case without altering the reservoir or replay
structure.
Parameterizing each ray segment by its endpoint coordinates
$\mathbf{p}_i = (\mathbf{u}_{i-1}, \mathbf{u}_i) \in \mathbb{R}^4$, a
forward-mode pass yields per-vertex Jacobians
$J_i = \partial \mathbf{p}_{i+1} / \partial \mathbf{p}_i$ that chain into a
path Jacobian $J_{0 \to i}$, mapping how a parameter perturbation propagates
along the path to move $u_i$, exactly the dependence the detached case
discards. We cache $J_{0 \to i}$ at constant extra storage when the reservoir
selects a vertex.

The selected connection contributes splat Jacobians
$\partial u_I / \partial \mathbf{p}_I$ and
$\partial C_I / \partial \mathbf{p}_I$, composed with the cached
$J_{0 \to I}$. During replay these are projected into each vertex's local
frame through the inverse accumulated Jacobian, propagating the radiometric
adjoint $\delta_L$ and positional adjoint $\delta_u = \nabla_u \mathcal{L}$
into the scene parameters; we regularize the rank-deficient Jacobians at
perfectly diffuse vertices, where the inverse is otherwise ill-defined.

\section{Verification}
\label{sec:verification}

We implement ResLRB as a custom integrator in Mitsuba~3~\cite{Mitsuba3} on top of Dr.Jit~\cite{Jakob2020DrJit}, and evaluate it against naive reverse-mode AD (Mitsuba's \texttt{ptracer}) and detached PRB on an NVIDIA RTX~5080 (16\,GB) using the CUDA/OptiX backend.

\paragraph{Gradient correctness}
Figures~\ref{fig:teaser} and~\ref{fig:validation} compare gradients under naive AD and ResLRB: the two agree up to Monte Carlo noise, the single retained connection, reweighted by $W_k / w_I$, reproducing the full per-path contribution in expectation.

\begin{figure}[t]
  \centering
  \begin{subfigure}[t]{\linewidth}
    \centering
    \includegraphics[width=\linewidth]{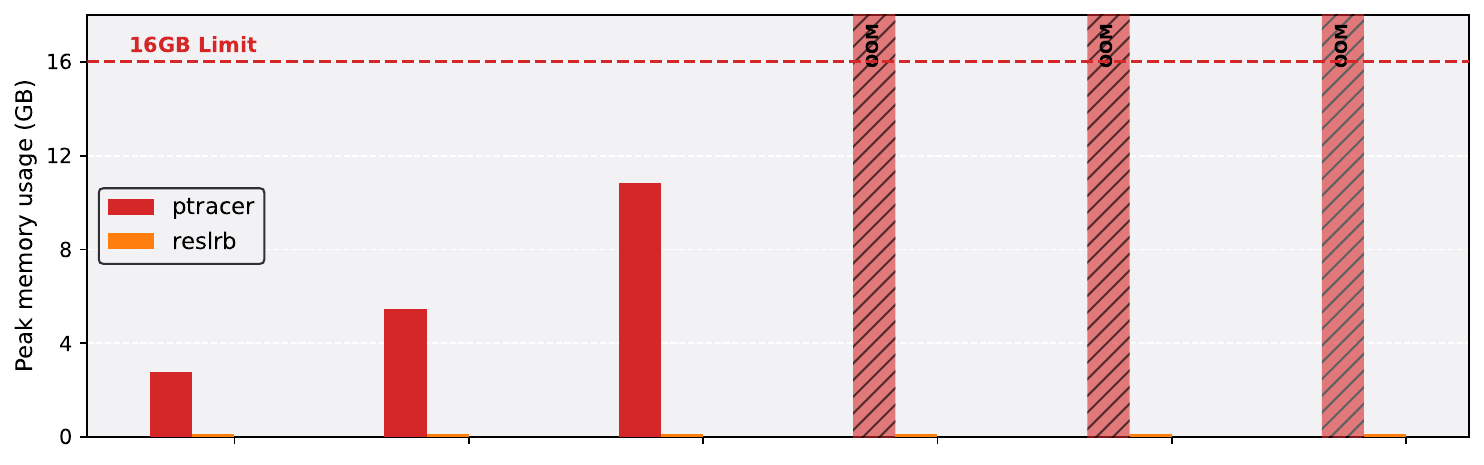}
    \label{fig:sweep-memory}
  \end{subfigure}
  \begin{subfigure}[t]{\linewidth}
    \centering
    \includegraphics[width=\linewidth]{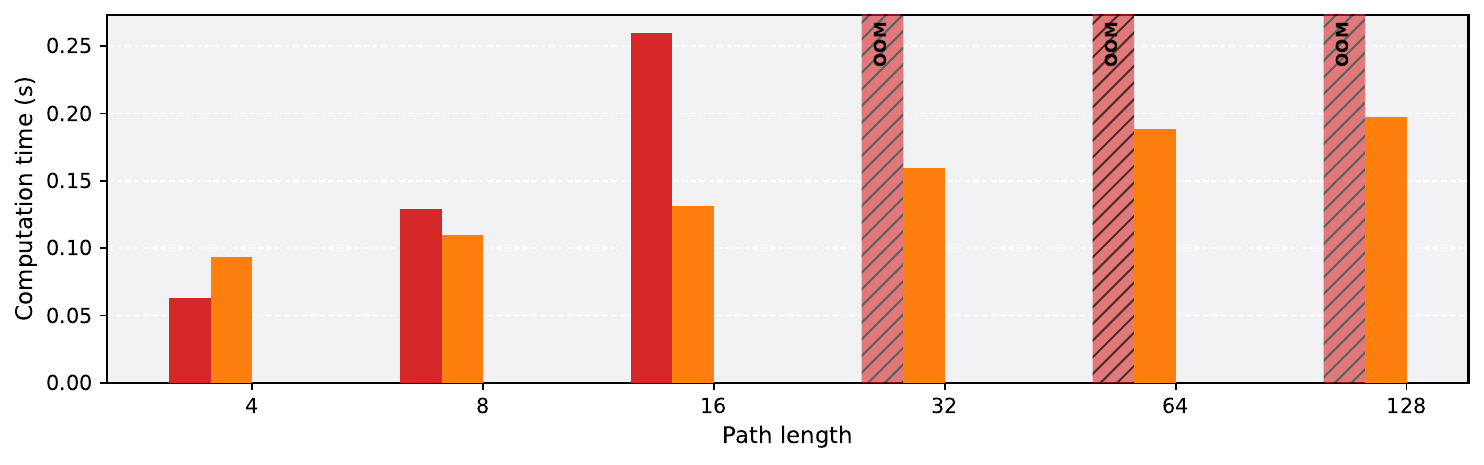}
    \label{fig:sweep-timing}
  \end{subfigure}
  \caption{Memory and time versus maximum path length. Naive AD (\texttt{ptracer}) grows linearly and exhausts memory past depth~16; ResLRB's bounded graph keeps memory flat and its per-step cost nearly so.}
  \Description{Two stacked line charts plotted against maximum path length on
the horizontal axis. In the upper chart, peak GPU memory on a logarithmic axis
rises linearly for naive automatic differentiation until its curve stops after
depth 16, where it exhausts memory, while the ResLRB curve stays flat at about
142 megabytes across the full range. In the lower chart, time per gradient step
shows naive automatic differentiation starting below ResLRB at depth 4, the two
crossing at depth 8, and naive automatic differentiation rising steeply to
roughly twice the ResLRB cost by depth 16; the ResLRB curve rises only
gradually, from 93 to 197 milliseconds out to depth 128.}
  \label{fig:sweep}
\end{figure}

\paragraph{Memory and timing}
Sweeping the maximum path depth (Figure~\ref{fig:sweep}), naive AD grows linearly and exhausts GPU memory beyond depth~16, whereas ResLRB holds memory constant at 142\,MB, bounded by the cost of differentiating a single scattering event. The replay traversal is what it pays for this: at depth~4 a gradient step costs $1.5\times$ that of naive AD (93 vs.\ 63\,ms), which needs only a single primal sweep. The overhead amortizes quickly as the memory traffic of the growing graph comes to dominate: the two are at parity by depth~8, and by depth~16 ResLRB is roughly twice as fast (132 vs.\ 260\,ms), beyond which naive AD no longer runs. From depth~4 to depth~128 ResLRB's own cost rises only from 93 to 197\,ms.

\section{Conclusion}
\label{sec:conclusion}

We described a two-pass adjoint formulation for differentiable light
tracing whose memory is constant in path length and in the number of
sensor connections per path, obtained by compressing the per-path
connections into a single-element weighted reservoir, replaying the path
deterministically during the adjoint pass, and opening a local AD scope
per vertex. Reservoirs of capacity $m > 1$ interpolate along the
memory--variance trade-off, multi-pass gathering schemes could retain every
connection at the cost of further traversals, and the attached extension
sketched above opens the method to geometric inverse problems. We expect
each of these to build naturally on the framework described here.

\begin{acks}
We acknowledge with gratitude the funding provided for this research by Simon Fraser University as well as a Discovery Grant provided by the Natural Sciences and Engineering Research Council of Canada. Additionally, we thank Benedikt Bitterli \cite{resources16} for providing The Grey \& White Room (from user Wig42 from blendswap).
\end{acks}

\bibliographystyle{ACM-Reference-Format}
\bibliography{references}

\end{document}